\documentclass[preprintnumbers,eqsecnum,prd,twocolumn]{revtex4}
\usepackage{graphicx}
\usepackage{latexsym}
\usepackage{amsmath}
\usepackage{amssymb}

\def\beq{\begin{equation}}
\def\eeq{\end{equation}}

\begin{document}

\title{Gravitational self-force corrections to tidal invariants for particles on circular orbits
in a  Kerr spacetime}

\author{Donato Bini}
\affiliation{
Istituto per le Applicazioni del Calcolo ``M. Picone,'' CNR, I-00185 Rome,
Italy}

\author{Andrea Geralico}
\affiliation{
Istituto per le Applicazioni del Calcolo ``M. Picone,'' CNR, I-00185 Rome,
Italy}

\date{\today}

\begin{abstract}
We generalize to the Kerr spacetime existing self-force results on tidal invariants for particles moving along circular orbits around a Schwarzschild black hole.
We obtain linear-in-mass-ratio corrections to the quadratic and cubic electric-type invariants and the quadratic magnetic-type invariant in series of the rotation parameter up to the fourth order. 
We then construct the eigenvalues of both electric and magnetic tidal tensors and analytically compute them through high post-Newtonian orders.
\end{abstract}

\maketitle

\section{Introduction}

Consider the general relativistic description of a two-body system in the case in which one body is spinning (say, with mass $m_2$ and spin $S_2=m_2a_2$) and the other is nonspinning, with mass (say $m_1$) much smaller than that of the companion, i.e., $m_1\ll m_2$.
In this situation, the small body affects the gravitational field generated by the large body (which can be identified with a background Kerr spacetime with parameters $m_2$ and $a_2$) by introducing a first-order perturbation proportional to the mass ratio $q=m_1/m_2$ of the two bodies,  
conveniently studied by using the Teukolsky formalism. With the aid of standard techniques one obtains the full perturbed metric, which is then suitably regularized and reconstructed along the world line of the small body and used to compute gauge-invariant quantities associated with physical observables.

Up to now it has been possible to compute in a Kerr spacetime with high accuracy (i.e., to a high post-Newtonian (PN) order) the corrections to 
gauge-invariant quantities which are continuous across the world line of the small body, namely its gravitational redshift in the field of the companion both in the case of circular and eccentric equatorial orbits \cite{Shah:2012gu,shah_MG14,Bini:2015xua,Bini:2016dvs,Kavanagh:2016idg}.
Using instead the Schwarzschild spacetime as a background, a number of gauge-invariant quantities generally discontinuous across the particle's world line has already been analytically computed, including the precession rate of a test gyroscope and the tidal curvature invariants (i.e., the tensorial contraction of the electric and magnetic parts of the Riemann tensor associated with natural observers) \cite{Bini:2012gu,Dolan:2013roa,Bini:2014ica,Bini:2014zxa,Dolan:2014pja,Kavanagh:2015lva,Shah:2015nva,Nolan:2015vpa,Kavanagh:2017wot,Bini:2018aps}.

Another natural description of the  two-body problem is the Hamiltonian one, where one solves the equations of motion in PN sense, order by order, starting from the flat background  where the two bodies live. Their mutual interaction generates dynamical corrections to the associated gravitational potentials, equivalent to curvature effects in perturbation theory. 
There exists a direct correspondence between these two points of view, so that new results from black hole perturbations can be converted in the Hamiltonian formalism (e.g., improving the knowledge of the gravitational potentials). Among the various coordinate-based Hamiltonian approaches (e.g., Arnowitt-Deser-Misner (ADM) coordinates and Harmonic coordinates), the so called \lq\lq effective one-body" (EOB) model \cite{Buonanno:1998gg,Buonanno:2000ef} --in short, a properly partially PN-resummed Hamiltonian model-- has proven to be very efficient (and incredibly fast in comparison with full numerical relativity simulations)  in 
following all the dynamics of the two bodies up to their merging. More the 250 thousands of EOB-based waveform templates have been generated in the analysis of the recently discovered gravitational wave signals by the LIGO and VIRGO detectors \cite{Abbott:2016blz,Abbott:2016nmj,Abbott:2017oio,TheLIGOScientific:2017qsa}. So far, translating into EOB new results from Kerr perturbations has been an important contribution of GSF in the last few years, although mainly orbital effects have been taken into account. The presence of spin in fact requires some care in the modeling of the \lq\lq effective" interaction and different EOB models with spin exist in the literature \cite{Damour:2014sva,Bohe:2016gbl}. 

The contribution of the present work is the analytical computation of linear-in-mass-ratio corrections to the quadratic and cubic electric-type and the quadratic magnetic-type tidal invariants for particles moving along circular equatorial orbits in a Kerr spacetime, generalizing previous results for a non-spinning black hole \cite{Bini:2012gu,Bini:2014zxa,Dolan:2014pja,Kavanagh:2015lva,Shah:2015nva,Nolan:2015vpa}.
We will follow a a well established procedure based on Teukolsky formalism and metric completion in a radiation gauge (see, e.g., Refs. \cite{Shah:2012gu,Bini:2015xua,Kavanagh:2016idg} and references therein). 
Therefore, we will limit ourselves to provide the final result with a minimum of related details.
This work continues some recent achievements on tidal invariants of Refs. \cite{Bini:2018svh,Bini:2018kov}, where GSF corrections to them have been computed in the case of spinning bodies on circular orbits as well spinless particles on slightly eccentric orbits around a Schwarzschild black hole.

\section{Kerr metric, perturbations and tidal invariants}

The Kerr metric with signature $-2$ and parameters $m_2$ and $a_2=a$ (with $\hat a =a/m_2$ dimensionless) written in Boyer-Lindquist coordinates reads
\begin{eqnarray}
\label{kerrmet}
d\bar s^2&=&\bar g_{\alpha\beta}dx^\alpha dx^\beta\nonumber\\
&=&\left(1-\frac{2m_2r}{\Sigma}  \right) dt^2+\frac{4am_2r \sin^2\theta}{\Sigma}dtd\phi\nonumber\\
&-& \frac{\Sigma}{\Delta}dr^2-\Sigma d\theta^2\nonumber\\
&-&  \left( r^2+a^2+\frac{2m_2ra^2\sin^2\theta}{\Sigma} \right)\sin^2\theta d\phi^2\,,
\end{eqnarray}
where   
\beq
\Delta= r^2+a^2-2m_2r\,,\qquad 
\Sigma=r^2+a^2\cos^2\theta\,.
\eeq
Let the perturbation be associated with a spinless particle of mass $m_1$ moving along a circular equatorial geodesic orbit at $r=r_0$, with four velocity
\beq
u^\alpha=\Gamma k^\alpha\,,\qquad k=\partial_t +\zeta \partial_\phi\,.
\eeq
Here $\zeta$ is the  constant angular velocity and $\Gamma=u^t$ is a normalization factor ($u^\alpha u_\alpha=1$), whose unperturbed values (denoted by a bar) are the following
\beq
m_2\bar\zeta=\frac{u^{3/2}}{1+\hat a u^{3/2}}\,,\qquad \bar\Gamma =\frac{1+\hat a u^{3/2}}{\sqrt{1-3u +2\hat a u^{3/2}}}\,,
\eeq
where $u=m_2/r$ is the dimensionless inverse radius of the orbit.
The constant vector $k$ generates a helical symmetry which is assumed to be a property of the (regularized) perturbed spacetime 
\beq
\label{pertmet}
g^{\rm R}_{\alpha\beta}=\bar g_{\alpha\beta}+q \, h^{\rm R}_{\alpha\beta} + O(q^2)\,,
\eeq
namely $k$ is supposed to be a Killing vector for the perturbed spacetime (it is also a Killing vector of the background spacetime, being a combination of $\partial_t$, generating time translations, and $\partial_\phi$, generating the rotational symmetry about the spin axis of the black hole).  

The particle is characterized by its energy-momentum tensor
\beq
T^{\alpha\beta}=\frac{m_1}{u^t r_0^2} u^\alpha u^\beta \delta_3\,,
\eeq
where
\beq
\delta_3=\delta(r-r_0)\delta \left(\theta-\frac{\pi}{2}\right)\delta (\phi-\bar\zeta t)\,,
\eeq
which enters the Teukolsky equation for the perturbed spin-weight $s=2$ Weyl scalar $\psi_0$ as a source term.
Decomposing $\psi_0$ in spheroidal angular harmonics and using the separability property of the Teukolsky equation in the frequency domain one ends up with a single radial equation to be solved in PN sense. From $\psi_0$ one then constructs the spin-2 Hertz potential $\Psi$ following a procedure due to Chrzanowski-Cohen-Kegeles and, eventually (by applying a proper second-order differential operator) one obtains the perturbed metric $h_{\alpha\beta}$ in the radiation gauge. All these steps are well established in the literature (see, e.g., Ref. \cite{Shah:2012gu} and references therein).
Finally, upon regularization, with the reconstructed perturbed metric and the four velocity field of the particle $u$, one can compute the perturbed Riemann tensor with associated electric and magnetic parts. The latters play the role of tidal potentials as explained below.

The geodesic condition in the perturbed spacetime implies
\begin{eqnarray}
m_2\zeta &=& \frac{u^{3/2}}{1+{\hat a}u^{3/2}}\left(1-q\frac{1+{\hat a}u^{3/2}}{4u^2} m_2[\partial_r h_{kk}]_1\right) \,,
\end{eqnarray}
with $h_{kk}=[h_{\mu\nu}^{\rm R}(x)k^\mu k^\nu]_1$, evaluated along the world line of the perturbing body.
Introducing the dimensionless frequency parameter
\beq
y=(m_2\zeta)^{2/3}\,,
\eeq
the previous equation can then be inverted to give
\beq
u = \frac{y}{(1-\hat ay^{3/2})^{2/3}}\left(1+q\frac{m_2[\partial_r h_{kk}]_1}{6y^2(1-\hat ay^{3/2})^{2/3}}\right)\,,
\eeq
which is used to reexpress the radius of the orbit in terms of a gauge-invariant variable.

\subsection{Tidal invariants}

We briefly recall below how to define the tidal potentials for a system of $N$ gravitationally interacting bodies through an effective action approach, as discussed in Ref. \cite{Bini:2012gu}. 
For point-mass objects with four velocities $u_A^\alpha=dx_A^\alpha /d\tau_A$ such a description is performed in terms of the action 
\beq
\label{eq:2.1}
S_0=\frac{c^4}{16 \pi G}\int \frac{d^4 x}{c}\sqrt{-g} R-\sum_A \int m_A c^2 d\tau_A\,,
\eeq
where $d\tau_A=-(u_A)_\mu dx_A^\mu$ is the regularized proper time along the world line $x_A^\mu(\tau_A)$ of body $A$. 
For extended objects, we have that each body feels the gravitational field of the whole system \cite{Damour:1982wm,Zhang86,Damour:1990pi,Damour:1991yw}, undergoing tidal effects which can be computed by adding other non-minimal couplings to $S_0$, involving higher-order derivatives of the field evaluated along the world line of each body (see, e.g., Ref. \cite{Levi:2015msa} and references therein).
The latter can be expressed in terms of the gravitoelectric ($G_L^A(\tau_A)\equiv G^A_{a_1\ldots a_l}(\tau_A)$) and gravitomagnetic ($H_L^A(\tau_A)\equiv H^A_{a_1\ldots a_l}(\tau_A)$) tidal tensors associated with the body $A$, which are symmetric and trace-free. 

The most general world line non-minimal action has then the form \cite{Damour:1990pi,Damour:1991yw,Damour:1992qi,Damour:1993zn,Bini:2012gu,Levi:2015msa}
\beq
S_{\rm non-min}  =\sum_{A}S_{\rm non-min} ^{A}\,,
\eeq
with
\begin{eqnarray}
\label{eq:2.2}
S_{\rm non-min} ^{A}
&= &  \frac{1}{4} \, \mu_A^{(2)} \int
d\tau_A \, G_{\alpha\beta}^A \, G_A^{\alpha\beta}\nonumber\\
&+&\frac{1}{6 \, c^2} \, \sigma_A^{(2)} \int d\tau_A \, H_{\alpha\beta}^A \,
H_A^{\alpha\beta}
+\ldots  \,.
\end{eqnarray}
We will consider here only the invariants associated with the quadrupolar electric-type and magnetic-type tidal tensors $G_{ab}$, $H_{ab}$,  related as follows to the spatial components of the \lq\lq electric" and \lq\lq magnetic" parts of the Riemann tensor 
\begin{eqnarray}
\label{eq:2.4}
G^A_{\alpha\beta}&\equiv & -{\mathcal E}(u_A)_{\alpha\beta}\,,\quad
H^A_{\alpha\beta} \equiv  2\, c\, {\mathcal B}(u_A)_{\alpha\beta}\,,
\end{eqnarray} 
where ${\mathcal E}(u_A)_{\alpha\beta}$ and ${\mathcal B}(u_A)_{\alpha\beta}$ are defined as
\begin{eqnarray}
\label{riemann_em}
{\mathcal E}(u_A)_{\alpha\beta}&=& R_{\alpha\mu\beta\nu}u_A^\mu u_A^\nu\,,\nonumber\\
{\mathcal B}(u_A)_{\alpha\beta}&=& [R^*]_{\alpha\mu\beta\nu}u_A^\mu u_A^\nu\,,
\end{eqnarray}
the symbol $^*$ denoting the spacetime dual of a tensor.
The associated non-minimal world line action (\ref{eq:2.2}) of the body 1 then reads
\begin{eqnarray}
\label{eq:2.2n}
S_{1\rm non-min}
&= &  \frac{1}{4} \, \mu_1^{(2)} \int d\tau_1 \, {\rm Tr}\,[{\mathcal E}(u_1)]^2\nonumber\\
&+& \frac{2}{3} \, \sigma_1^{(2)} \int d\tau_1\, {\rm Tr}\,[{\mathcal B}(u_1)]^2
+\ldots  \,,  
\end{eqnarray}
where we have set $G=c=1$.
Hereafter, we will omit the body label $A=1$ to  ease  notation.

In the present paper we compute first-order GSF corrections to the quadratic tidal-electric and tidal-magnetic invariants ${\rm Tr}\,[{\mathcal E}(u)]^2$ and ${\rm Tr}\,[{\mathcal B}(u)]^2$ as well as to the cubic tidal-electric invariant ${\rm Tr}\,[{\mathcal E}(u)]^3$, and evaluate the eigenvalues of the tidal tensors ${\mathcal E}(u)$ and ${\mathcal B}(u)$.
For convenience, we will work with their rescaled counterparts
\beq
{\mathcal E}(u)=\Gamma^2 {\mathcal E}(k)\,,\qquad {\mathcal B}(u)=\Gamma^2 {\mathcal B}(k)\,,
\eeq
with associated invariants
\begin{eqnarray}
\label{invar}
{\mathcal J}_{e^2}&\equiv&m_2^4{\rm Tr}\,[{\mathcal E}(k)]^2\,,\nonumber\\
{\mathcal J}_{b^2}&\equiv&m_2^4{\rm Tr}\,[{\mathcal B}(k)]^2\,,\nonumber\\
{\mathcal J}_{e^3}&\equiv&m_2^6{\rm Tr}\,[{\mathcal E}(k)]^3\,.
\end{eqnarray}

\section{GSF computation of tidal invariants}

The first-order self-force (1SF) accurate expansions of the electric-type and magnetic-type tidal invariants \eqref{invar} read
\begin{eqnarray}
{\mathcal J}_{e^2}&=&
{\mathcal J}_{e^2}^{(0)}[1+q\, \delta_{e^2}(y)]+O(q^2)\,,\nonumber\\
{\mathcal J}_{b^2}&=&
{\mathcal J}_{b^2}^{(0)}
[1+q \delta_{b^2}(y)]+O(q^2)\,, \nonumber\\
{\mathcal J}_{e^3}&=&
{\mathcal J}_{e^3}^{(0)}
[1+q \delta_{e^3}(y)] +O(q^2)\,,
\end{eqnarray}
where 
\begin{widetext}
\begin{eqnarray}
{\mathcal J}_{e^2}^{(0)}&=&
6u^6\frac{1-3u+3u^2-2{\hat a}u^{3/2}(1+3{\hat a}^2u^2)+3{\hat a}^2u^2(1+{\hat a}^2u^2)+{\hat a}^2u^3}{1+{\hat a}u^{3/2}}
\,,\nonumber\\
{\mathcal J}_{b^2}^{(0)}&=&
18u^7(1-2u+{\hat a}^2u^2)\frac{(1-{\hat a}u^{1/2})^2}{(1+{\hat a}u^{3/2})^4} 
\,, \nonumber\\
{\mathcal J}_{e^3}^{(0)}&=&
-3u^9(1-3u+2{\hat a}u^{3/2})(2-3u-2{\hat a}u^{3/2}+3{\hat a}^2u^2)\frac{1-4{\hat a}u^{3/2}+3{\hat a}^2u^2}{(1+{\hat a}u^{3/2})^6}
\,,
\end{eqnarray}
\end{widetext}
denote the corresponding unperturbed values, with $u = {y}/{(1-{\hat a}y^{3/2})^{2/3}}$.

The expressions of the first order corrections $\delta_{e^2}$, $\delta_{b^2}$ and $\delta_{e^3}$ in terms of the components of the perturbed metric and their derivatives are listed in Appendix.
Their regularized values are given by the convergent series \cite{Bini:2014zxa} 
\beq
\delta^{\rm reg}=\sum_{l=0}^\infty \left( \delta_l^0-B(y;l)\right)\,,
\eeq
where 
\beq
\delta_l^0\equiv \frac12 (\delta_l^+ + \delta_l^-)\,,
\eeq
and the \lq\lq subtraction term'' is of the form
\beq
B(y;l)=l(l+1)b_0(y)+b_1(y)\,,
\eeq
with $b_0(y)=b_0^{a^0}(y)+{\hat a} b_0^{a^1}(y)+{\hat a}^2 b_0^{a^2}(y)+{\hat a}^3 b_0^{a^3}(y)+{\hat a}^4 b_0^{a^4}(y)$ and similarly for $b_1(y)$.
It is known that the regularization procedure comprises several subtleties related to the fact that one should actually subtract the full Detweiler-Whiting singular field (see, e.g., Ref. \cite{Heffernan:2012su} and references therein), implying in general the existence of various regularization parameters ($A$, $B$, $C$, $D$, etc.). We take left-right averages across the particle's world line and only subtract the $B$-term, which is enough to have a convergent series. 
The completion of the metric then requires the contribution of nonradiative multipoles $l=0,1$, which has been recently obtained in Ref. \cite{mvdm}.

We will omit showing explicitly the final results for $\delta_{e^2}$, $\delta_{b^2}$ and $\delta_{e^3}$ after regularization, focusing only on the associated eigenvalues of the tidal tensors.

\subsection{Eigenvalues}

The eigenvalues of the tidal-electric and tidal-magnetic quadrupolar tensors $m_2^2{\mathcal E}(u)^\mu{}_\nu$ and  $m_2^2{\mathcal B}(u)^\mu{}_\nu$ are  such that 
\begin{eqnarray}
\label{eq:4.1}
m_2^2 {\mathcal E}(u)&=& {\rm diag} [\lambda_1^{\rm (E)},\lambda_2^{\rm (E)},-(\lambda_1^{\rm (E)}+\lambda_2^{\rm (E)})]\,,\nonumber\\
m_2^2 {\mathcal B}(u)&=& {\rm diag} [\lambda^{\rm (B)},-\lambda^{\rm (B)},0]\,,
\end{eqnarray}
where we have used their traceless property, and the existence of a zero eigenvalue of ${\mathcal B}(u)$ \cite{Dolan:2014pja}. 
They are related to the eigenvalues $\sigma_a^{\rm (E)}$ and $\sigma^{\rm (B)}$ of the rescaled tidal tensors ${\mathcal E}(k)$ and ${\mathcal B}(k)$ by 
\begin{eqnarray}
\label{eq:4.5}
\lambda_a^{\rm (E)}=\Gamma^2 \sigma_a^{\rm (E)}\,,\qquad
\lambda^{\rm (B)}=\Gamma^2 \sigma^{\rm (B)}\,,
\end{eqnarray}
where 
\beq
\Gamma= \frac{1+{\hat a}u^{3/2}}{\sqrt{1-3u+2{\hat a}u^{3/2}}}+q\,\delta U(y)\,,
\eeq
with $u = {y}/{(1-{\hat a}y^{3/2})^{2/3}}$ and the 1SF expansion $\delta U(y)$ has been derived in our previous work \cite{Bini:2015xua}.

As usual, we will write
\begin{eqnarray}
\lambda_a^{\rm (E)}&=&\lambda_a^{\rm (E)\,0SF}+q\lambda_a^{\rm(E)\,1SF}\,,\nonumber\\
\lambda^{\rm (B)}&=&\lambda^{\rm (B)\,0SF}+q\lambda^{\rm(B)\,1SF}\,,
\end{eqnarray}
where the unperturbed (0SF) values are
\begin{eqnarray}
\label{eq:4.3}
\lambda_1^{\rm (E)\,0SF}&=& -u^3 \frac{2-3u-2{\hat a}u^{3/2}+3{\hat a}^2u^2}{1-3u+2{\hat a}u^{3/2}}\,,\nonumber\\
\lambda_2^{\rm (E)\,0SF}&=& u^3 \frac{1-4{\hat a}u^{3/2}+3{\hat a}^2u^2}{1-3u+2{\hat a}u^{3/2}}\,,\nonumber\\
\lambda^{\rm (E)\,0SF}&=& 3u^{7/2}(1-{\hat a}u^{1/2})\frac{\sqrt{1-2u+{\hat a}^2u^2}}{1-3u+2{\hat a}u^{3/2}} \,,
\end{eqnarray}
with $u = {y}/{(1-{\hat a}y^{3/2})^{2/3}}$.
The 1SF corrections to the rescaled eigenvalues $\sigma_a^{\rm (E)}$ and $\sigma^{\rm (B)}$ are computed following Ref. \cite{Bini:2014zxa}, so that one 1PN level in the analytic accuracy of $\sigma_2^{\rm (E)}$ is lost. 

We also use the notation
\begin{eqnarray}
\lambda_1^{\rm(E)\,1SF}(y)&=&\lambda_1^{\rm(E)\,1SF\,a^0}(y)+{\hat a} \lambda_1^{\rm(E)\,1SF\,a^1}(y)\nonumber\\
&+&{\hat a}^2 \lambda_1^{\rm(E)\,1SF\,a^2}(y)+{\hat a}^3 \lambda_1^{\rm(E)\,1SF\,a^3}(y)\nonumber\\
&+&{\hat a}^4 \lambda_1^{\rm(E)\,1SF\,a^4}(y)\,,
\end{eqnarray}
and similarly for the others.

The Schwarzschild values $\lambda^{\rm (E)\, 1SF}_{1 \, a^0}$, $\lambda^{\rm (E)\, 1SF}_{2 \, a^0}$ and $\lambda^{\rm (B)\, 1SF}_{a^0}$ are known with high PN accuracy \cite{Bini:2014zxa,Dolan:2014pja,Kavanagh:2015lva,Nolan:2015vpa}.
We recall below for completeness only the first few terms

\begin{widetext}

\begin{eqnarray}
\lambda^{\rm (E)\, 1SF}_{1 \, a^0} &=& 2 y^3+2 y^4-\frac{19}{4}y^5+\left(\frac{227}{3}-\frac{593}{256}\pi^2\right) y^6\nonumber\\
&&+\left(-\frac{71779}{4800}-\frac{719}{256}\pi^2+\frac{1536}{5}\ln(2)+\frac{384}{5}\ln(y)+\frac{768}{5}\gamma\right)y^7
+O_{{\rm ln}}(y^8)
\,,\nonumber\\
\lambda^{\rm (E)\, 1SF}_{2 \, a^0}&=&  -y^3-\frac32 y^4-\frac{23}{8}y^5+\left(-\frac{2593}{48}+\frac{1249}{1024}\pi^2\right) y^6\nonumber\\
&& +\left(-\frac{362051}{3200}-\frac{128}{5}\ln(y)+\frac{1737}{1024}\pi^2-\frac{256}{5}\gamma-\frac{512}{5}\ln(2)\right) y^7
+O_{{\rm ln}}(y^8)
\,,\nonumber\\
-\lambda^{\rm (B)\, 1SF}_{a^0}&=&  
2y^{7/2}+3y^{9/2}+\frac{59}{4}y^{11/2}+\left(\frac{2761}{24}-\frac{41}{16}\pi^2\right)y^{13/2}\nonumber\\
&&+\left(\frac{1618039}{2880}-\frac{112919}{3072}\pi^2+\frac{1808}{15}\gamma+240\ln(2)+\frac{904}{15}\ln(y) \right)y^{15/2}
+O_{{\rm ln}}(y^{17/2})
\,.
\end{eqnarray}

The $O(\hat a^1)$--$O(\hat a^4)$ contributions for each eigenvalue are the main original contribution of the present work and are listed below.

Results for $\lambda_1^{\rm(E)\,1SF}$:

\begin{eqnarray}
\lambda_1^{\rm(E)\,1SF\,a^1}(y)&=&
-4 y^{9/2}-\frac{95}{3} y^{11/2}-\frac{923}{6} y^{13/2}\nonumber\\
&+&\left(-\frac{44357}{72}+\frac{1165}{768}\pi^2\right) y^{15/2}\nonumber\\
&+&\left(-\frac{23584579}{7200}+\frac{119515}{1536}\pi^2-\frac{4784}{15}\gamma-\frac{2392}{15}\ln(y)-\frac{9392}{15}\ln(2)\right) y^{17/2}\nonumber\\
&+&\left(-\frac{22210361969}{4233600}-\frac{693016}{315}\gamma-\frac{358604}{315}\ln(y)-\frac{325576}{105}\ln(2)-\frac{20451899}{147456}\pi^2-\frac{9234}{7}\ln(3)\right) y^{19/2}\nonumber\\
&-& \frac{825032}{1575}\pi y^{10}\nonumber\\
&+&\left(-\frac{58928132968141}{228614400}+\frac{1374011}{2835}\ln(y)+\frac{3945526}{2835}\gamma-\frac{15750506}{2835}\ln(2)+\frac{21870}{7}\ln(3)\right. \nonumber\\
&&\left. +\frac{10601643869}{589824}\pi^2+\frac{250702133}{524288}\pi^4\right) y^{21/2}\nonumber\\
&-& \frac{3849604}{2205}\pi y^{11}\nonumber\\
&+&\left(-\frac{38054676977}{32744250}\ln(y)-\frac{18546721509346775267}{19363639680000}+\frac{1640648}{1575}\ln(y)^2+\frac{4568981023}{16372125}\gamma\right.\nonumber\\
&& +\frac{5235008}{315}\gamma\ln(2)-\frac{74432}{15}\zeta(3) + \frac{19219584847}{654885}\ln(2)\nonumber\\
&+&\frac{224125947}{12320}\ln(3)-\frac{166015625}{9504}\ln(5)+\frac{1694532559908697}{11890851840}\pi^2-\frac{2226578089167}{335544320}\pi^4+\frac{26137376}{1575}\ln(2)^2\nonumber\\
&+&\left. \frac{6562592}{1575}\gamma^2+\frac{6562592}{1575}\gamma\ln(y)+\frac{2617504}{315}\ln(2)\ln(y)\right) y^{23/2}\nonumber\\
&+&\frac{24229293377}{6548850}\pi y^{12}
+O_{\rm ln}(y^{25/2})\,,\nonumber
\end{eqnarray}
\begin{eqnarray}
\lambda_1^{\rm(E)\,1SF\,a^2}(y)&=&
 5 y^5+\frac{53}{2} y^6+\frac{9539}{72} y^7+\left(\frac{131101}{144}-\frac{3343}{512}\pi^2\right) y^8\nonumber\\
&+&\left(\frac{13469583}{3200}-\frac{292797}{8192}\pi^2+\frac{4112}{5}\ln(2)+\frac{1032}{5}\ln(y)+\frac{2064}{5}\gamma\right) y^9\nonumber\\
&+&\left(\frac{151504}{315}\ln(y)+\frac{9439449583}{1209600}+\frac{303008}{315}\gamma+\frac{161312}{315}\ln(2)+\frac{9477}{7}\ln(3)+\frac{1053261163}{1179648}\pi^2\right) y^{10}\nonumber\\
&+& \frac{219992}{525}\pi y^{21/2}\nonumber\\
&+&\left(\frac{3779243}{2835}\ln(y)+\frac{29633535218351}{101606400}+\frac{529502}{567}\gamma+\frac{4608}{5}\zeta(3)+\frac{36439126}{2835}\ln(2)-\frac{79623}{14}\ln(3)\right. \nonumber\\
&& \left.-\frac{21842364000733}{1101004800}\pi^2-\frac{7984116587}{67108864}\pi^4\right) y^{11}\nonumber\\
&+& \frac{15603907}{11025}\pi y^{23/2}\nonumber\\
&+&
\left(\frac{24488109508}{606375}\gamma-\frac{280768}{21}\gamma\ln(2)+\frac{1323559510617011}{64424509440}\pi^4+\frac{1986757084}{72765}\ln(2)+\frac{289495377}{24640}\ln(3)\right.\nonumber\\
&&
-\frac{585504}{175}\gamma^2-\frac{2338592}{175}\ln(2)^2+\frac{283203125}{19008}\ln(5)+\frac{55872}{5}\zeta(3)+\frac{25199471184842137}{79272345600}\pi^2\nonumber\\
&&
+\frac{14829961154}{606375}\ln(y)-\frac{585504}{175}\gamma\ln(y)-\frac{140384}{21}\ln(2)\ln(y)-\frac{146376}{175}\ln(y)^2\nonumber\\
&&\left.
-\frac{16911655702097901527}{3520661760000}\right)y^{12}
+O_{\rm ln}(y^{25/2})
\,,\nonumber
\end{eqnarray}
\begin{eqnarray}
\lambda_1^{\rm(E)\,1SF\,a^3}(y)&=&
 \frac53 y^{13/2}-55 y^{15/2}-\frac{324305}{648} y^{17/2}\nonumber\\
&+&\left(-\frac{10722523}{4050}-\frac{34019}{1536}\pi^2-\frac{768}{5}\zeta(3)-\frac{576}{5}\ln(y)-\frac{576}{5}\ln(2)\right) y^{19/2}\nonumber\\
&+&\left(-\frac{854990743}{28800}+\frac{60387577}{49152}\pi^2-608\ln(y)-\frac{704}{5}\gamma-768\zeta(3)-\frac{3904}{5}\ln(2)\right) y^{21/2}\nonumber\\
&+&\left(-\frac{2142752}{315}\ln(y)-\frac{2235738608977}{7620480}-\frac{2255776}{315}\gamma-\frac{23296}{5}\zeta(3)-\frac{2043824}{135}\ln(2)\right.\nonumber\\
&&\left.-\frac{15552}{7}\ln(3)+\frac{3535975103}{165888}\pi^2\right) y^{23/2}\nonumber\\
&-&\frac{203368}{175}\pi y^{12}
+O_{\rm ln}(y^{25/2})
\,,\nonumber 
\end{eqnarray}
\begin{eqnarray}
\lambda_1^{\rm(E)\,1SF\,a^4}(y)&=&
\frac{263}{18} y^8+91 y^9+\left(\frac{3868903}{3888}+\frac{13215}{8192}\pi^2\right) y^{10}\nonumber\\
&+&\left(\frac{75631214}{8505}+\frac{72077881}{9830400}\pi^2-60\ln(y)+264\gamma+\frac{90368}{15}\zeta(3)+328\ln(2)-6144\zeta(5)-\frac{54784}{7875}\pi^4\right) y^{11}\nonumber\\
&+&\left(\frac{75764}{35}\gamma-\frac{496448}{7875}\pi^4+\frac{146956}{105}\ln(2)+1458\ln(3)+\frac{434432}{15}\zeta(3)-29952\zeta(5)\right.\nonumber\\
&&\left.
+\frac{22240271822549}{550502400}\pi^2-\frac{319903684889}{940800}-\frac{10614}{35}\ln(y)\right)y^{12}+O_{\rm ln}(y^{25/2})
\,.
\end{eqnarray}

Results for $\lambda_2^{\rm(E)\,1SF}$:

\begin{eqnarray}
\lambda_2^{\rm(E)\,1SF\,a^1}(y)&=&
5 y^{9/2}+26 y^{11/2}+\frac{3451}{24} y^{13/2}\nonumber\\
&&+\left(\frac{49687}{72}-\frac{2495}{768}\pi^2\right) y^{15/2}\nonumber\\
&&+\left(\frac{47694017}{9600}-\frac{975049}{4096}\pi^2+\frac{2096}{15}\ln(y)+\frac{8288}{15}\ln(2)+\frac{4192}{15}\gamma \right) y^{17/2}\nonumber\\
&&+\left(\frac{397832}{315}\ln(2)+\frac{53068}{105}\ln(y)+\frac{34467328501}{352800}+\frac{103448}{105}\gamma-\frac{220232047}{24576}\pi^2+\frac{4860}{7}\ln(3)\right) y^{19/2}\nonumber\\
&&+\frac{78704}{225}\pi y^{10}\nonumber\\
&&+\left(\frac{154324193851141}{182891520}-\frac{41768}{81}\ln(y)-\frac{500624}{405}\gamma+\frac{3626512}{2835}\ln(2)-\frac{13122}{7}\ln(3)\right.\nonumber\\
&& \left.-\frac{421991832085}{4718592}\pi^2+\frac{1798084955}{2097152}\pi^4\right) y^{21/2}\nonumber\\
&&+\frac{6246043}{7350}\pi y^{11}
+O_{\rm ln}(y^{23/2})\,,\nonumber 
\end{eqnarray}
\begin{eqnarray}
\lambda_2^{\rm(E)\,1SF\,a^2}(y)&=&
-5 y^5-21 y^6-\frac{3647}{24} y^7+\left(-\frac{34247}{36}+\frac{12427}{2048}\pi^2\right) y^8\nonumber\\
&&+\left(-\frac{40143929}{9600}+\frac{582859}{65536}\pi^2-\frac{772}{5}\ln(y)-\frac{1544}{5}\gamma-616\ln(2)\right) y^9\nonumber\\
&&+\left(-\frac{68134}{315}\ln(y)+\frac{719136473}{8064}-\frac{136268}{315}\gamma-\frac{25804}{315}\ln(2)-729\ln(3)-\frac{8635481489}{786432}\pi^2\right) y^{10}\nonumber\\
&&-\frac{4708}{15}\pi y^{21/2}\nonumber\\
&&+\left(-\frac{1144565}{567}\gamma-\frac{42961473011}{536870912}\pi^4-\frac{24381769}{2835}\ln(2)+\frac{17820}{7}\ln(3)-\frac{1536}{5}\zeta(3)\right.\nonumber\\
&&\left.
-\frac{12863257246761179}{39636172800}\pi^2-\frac{1531637}{1134}\ln(y)+\frac{63382286411599}{20321280}\right)y^{11}+O_{\rm ln}(y^{23/2})
\,,\nonumber
\end{eqnarray}
\begin{eqnarray}
\lambda_2^{\rm(E)\,1SF\,a^3}(y)&=&
-\frac53  y^{13/2}+62 y^{15/2}+\frac{100469}{216} y^{17/2}\nonumber\\
&&+\left(\frac{22598287}{8100}+\frac{192}{5}\ln(y)+\frac{256}{5}\zeta(3)+\frac{192}{5}\ln(2)+\frac{283177}{12288}\pi^2\right) y^{19/2}\nonumber\\
&&+\left(\frac{278665423}{5760}+\frac{1536}{5}\ln(y)+240\gamma+\frac{1568}{5}\zeta(3)+632\ln(2)-\frac{49870393}{16384}\pi^2\right) y^{21/2}\nonumber\\
&&+O_{\rm ln}(y^{23/2})
\,,\nonumber
\end{eqnarray}
\begin{eqnarray}
\lambda_2^{\rm(E)\,1SF\,a^4}(y)&=&
 -\frac{263}{18} y^8-\frac{625}{8} y^9+\left(-\frac{683999}{648}-\frac{135495}{65536}\pi^2\right) y^{10}\nonumber\\
&&+\left(-\frac{24615820919}{2721600}-264\gamma+\frac{54784}{23625}\pi^4-\frac{2216}{5}\ln(2)-\frac{31232}{15}\zeta(3)+2048\zeta(5)+\frac{2919530863}{58982400}\pi^2\right.\nonumber\\
&&\left.
-\frac{276}{5}\ln(y)\right)y^{11}
+O_{\rm ln}(y^{23/2})
\,.
\end{eqnarray}

Results for $\lambda^{\rm(B)\,1SF}$:

\begin{eqnarray}
-\lambda^{\rm(B)\,1SF\,a^1}(y)&=&
-4 y^4-\frac{28}{3} y^5-\frac{121}{2} y^6+\left(-384+\frac{41}{8}\pi^2\right) y^7\nonumber\\
&&+\left(-\frac{1609523}{1440}+\frac{497}{192}\pi^2-\frac{516}{5}\ln(y)-\frac{1032}{5}\gamma-\frac{2056}{5}\ln(2)\right) y^8\nonumber\\
&& +\left(\frac{498497509}{50400}-\frac{763429}{512}\pi^2+\frac{9956}{21}\ln(2)+\frac{318}{35}\ln(y)+\frac{636}{35}\gamma-\frac{2916}{7}\ln(3)\right) y^9\nonumber\\
&& -\frac{109996}{525}\pi y^{19/2}\nonumber\\
&&+\left(\frac{223267}{2835}\gamma+\frac{21059401}{262144}\pi^4-\frac{217951}{81}\ln(2)+\frac{13365}{7}\ln(3)-\frac{3314946773}{442368}\pi^2\right.\nonumber\\
&&\left.
+\frac{1067165682679}{25401600}-\frac{30749}{5670}\ln(y)\right)y^{10}\nonumber\\
&&-\frac{971542}{11025}\pi y^{21/2}\nonumber\\
&&+\left(-\frac{586496084291}{32744250}\gamma+\frac{140384}{21}\gamma\ln(2)-\frac{603286332419}{201326592}\pi^4-\frac{6239060383}{261954}\ln(2)\right.\nonumber\\
&&
-\frac{24619869}{6160}\ln(3)+\frac{292752}{175}\gamma^2+\frac{1169296}{175}\ln(2)^2-\frac{9765625}{4752}\ln(5)-\frac{16416}{5}\zeta(3)\nonumber\\
&&
+\frac{58939924322641}{928972800}\pi^2+\frac{73188}{175}\ln(y)^2+\frac{292752}{175}\gamma\ln(y)+\frac{70192}{21}\ln(2)\ln(y)\nonumber\\
&&\left.
-\frac{88892510118308059}{220041360000}-\frac{605985461891}{65488500}\ln(y)\right)y^{11}\nonumber\\
&&+\frac{7620479}{66825}\pi y^{23/2}
+O_{\rm ln}(y^{12})
\,,\nonumber
\end{eqnarray}
\begin{eqnarray}
-\lambda^{\rm(B)\,1SF\,a^2}(y)&=&
 \frac{10}{3} y^{11/2}+\frac{604}{9} y^{13/2}+\frac{1457}{4} y^{15/2}+\left(\frac{64885}{36}+\frac{9631}{1024}\pi^2\right) y^{17/2}\nonumber\\
&&+\left(\frac{281701261}{14400}-\frac{12600593}{12288}\pi^2+\frac{976}{5}\ln(y)+\frac{1952}{5}\gamma+\frac{2272}{3}\ln(2)\right) y^{19/2}\nonumber\\
&&
+\left(\frac{147786}{35}\gamma+\frac{716126}{105}\ln(2)+\frac{44469}{28}\ln(3)-\frac{9838313339}{196608}\pi^2+\frac{226607}{105}\ln(y)+\frac{23409038221}{44100}\right)y^{21/2}\nonumber\\
&&+\frac{207652}{315}\pi y^{11}\nonumber\\
&&+\left(\frac{36534419}{8505}\gamma+\frac{199119651947}{50331648}\pi^4+\frac{124957859}{8505}\ln(2)-\frac{99387}{280}\ln(3)+\frac{3072}{5}\zeta(3)\right.\nonumber\\
&&\left.
-\frac{197722622556217}{928972800}\pi^2+\frac{48690899}{17010}\ln(y)+\frac{860650678007317}{457228800}\right)y^{23/2}+O_{\rm ln}(y^{12})
\,,\nonumber
\end{eqnarray}
\begin{eqnarray}
-\lambda^{\rm(B)\,1SF\,a^3}(y)&=&
-3 y^6-\frac{463}{18} y^7-\frac{96337}{648} y^8+\left(-\frac{20683}{16}+\frac{429}{128}\pi^2\right) y^9\nonumber\\
&&+\left(-\frac{2035430657}{259200}+\frac{1101311}{18432}\pi^2-\frac{1828}{5}\gamma-\frac{3812}{5}\ln(2)-\frac{512}{5}\zeta(3)-\frac{1106}{5}\ln(y)\right)y^{10}\nonumber\\
&&+\left(-\frac{25730}{21}\gamma-\frac{495566}{315}\ln(2)-\frac{8019}{7}\ln(3)-\frac{3776}{5}\zeta(3)-\frac{25089030215}{663552}\pi^2\right.\nonumber\\
&&\left.
-\frac{100949}{105}\ln(y)+\frac{729564502223}{2177280}\right)y^{11}
\nonumber\\
&&
-\frac{38734}{105}\pi y^{23/2}
+O_{\rm ln}(y^{12})
\,,\nonumber
\end{eqnarray}
\begin{eqnarray}
-\lambda^{\rm(B)\,1SF\,a^4}(y)&=&
-\frac{15}{4} y^{15/2}+\frac{10627}{648} y^{17/2}+\frac{2983073}{7776} y^{19/2}\nonumber\\
&&+\left(\frac{12732217}{4800}+\frac{384}{5}\ln(2)+\frac{2048}{15}\zeta(3)+\frac{32339}{2048}\pi^2+\frac{384}{5}\ln(y)\right)y^{21/2}\nonumber\\
&&+\left(-\frac{406}{3}\gamma-\frac{89024}{23625}\pi^4+\frac{298}{3}\ln(2)+\frac{11840}{3}\zeta(3)-3328\zeta(5)-\frac{4998352104167}{928972800}\pi^2\right.\nonumber\\
&&\left.
+\frac{5033}{15}\ln(y)+\frac{60509611189}{806400}\right)y^{23/2}
+O_{\rm ln}(y^{12})
\,.
\end{eqnarray}

\end{widetext}

In order to associate a theoretical error to our analytical expressions we follow the discussion of Ref. \cite{Bini:2014zxa} (see, e.g., Eq. (4.18) there).
The 1SF corrections to the eigenvalues are given as series expansion with respect to the black hole rotation parameter $\hat a$, so that we expect that each term generally diverges at the Schwarzschild light-ring $y=1/3$.
Therefore, we use the following estimate of the theoretical error \cite{Bini:2014zxa}
\beq
\sigma_N^{\rm th}(y)\simeq C_{N+\frac12} \frac{(3y)^{N+\frac12}}{(1-3y)^{a_N}}\,,
\eeq
where $N$ is the maximum power appearing in the PN expansion of each term $\lambda_i^{{\rm 1SF}\, a^n}$, and $a_N$ is an adjustable parameter which can be suitably chosen in order to improve the agreement with numerical data, if available. In this case no numerical data exist and hence we assume it to be zero.
Furthermore, the (positive) coefficients $C_{N+\frac12}$ are roughly of the order of unity, typically between 1 and 10.
For example, the estimated error on $\lambda_1^{{\rm (E)\, 1SF}\, a^1}$ at $y=0.1$ turns out to be $\sigma_{12}^{\rm th}(0.1)\approx 10^{-7}$, so that $\lambda_1^{{\rm (E)\, 1SF}\, a^1}(0.1)\approx -0.0003077$.
This estimate can be also checked by considering the various successive PN-approximants and identifying the digits which stabilize as the PN order increases, as summarized in Table 1.
Similar considerations hold for the other coefficients $\lambda_i^{{\rm 1SF}\, a^n}$.


\begin{table*}
\centering
\caption{A list of numerical values of $\lambda_1^{{\rm (E)\, 1SF}\, a^n}$ for $y=0.1$.
}
\begin{ruledtabular}
\begin{tabular}{|c|l|l|l|l|}
{\rm PN} &$ \lambda_1^{{\rm (E)\, 1SF}\, a^1}(0.1)$&$ \lambda_1^{{\rm (E)\, 1SF}\, a^2}(0.1)$&$ \lambda_1^{{\rm (E)\, 1SF}\, a^3}(0.1)$&$ \lambda_1^{{\rm (E)\, 1SF}\, a^4}(0.1)$\cr
\hline
7   & -0.0002942847 &  0.0000897486 & -0.0000012122 &  - \cr
8   & -0.0003030081 &  0.0000982084 & -0.0000027948 &  0.0000001461 \cr
9   & -0.0003058106 &  0.0001023980 & -0.0000037009 &  0.0000002371   \cr
10  & -0.0003070667 &  0.0001042302 & -0.0000042609 &  0.0000003382  \cr
11  & -0.0003077093 &  0.0001050993 & -0.0000045455 &  0.0000004350  \cr
12  & -0.0003076977 &  0.0001054684 & -0.0000045492 &  0.0000004958
\end{tabular}
\end{ruledtabular}
\label{tab:1}
\end{table*}

\section{Concluding remarks}

We have computed the first-order GSF corrections to both the electric-type and magnetic-type tidal eigenvalues for particles moving along circular orbits in a Kerr spacetime, generalizing previous results valid for the Schwarzschild case.
The computation is performed as a power series of the black hole rotation parameter $\hat a$ (up to $O(\hat a^4)$ included) and through a high-PN order in terms of the gauge-invariant frequency variable $y$.  

These results are ready to be converted into other formalisms, like the EOB model, entering the $S_1$ corrections to the tidal part of the Hamiltonian, eventually Pad\'e-resummed.
However, this would require a specific treatment which goes beyond the scopes of the present paper. We will leave it for future works.\\

\appendix

\section{Tidal invariants in a perturbed Kerr metric}

We list below the expressions of the rescaled tidal invariants in terms of the components of the perturbed metric and their derivatives.
All quantities are evaluated at $u = {y}/{(1-{\hat a}y^{3/2})^{2/3}}$.

\begin{widetext}

\subsection{Quadratic electric-type invariant ${\rm Tr}\,[{\mathcal E}(k)]^2$}

The $O(q)$ perturbation to $m_2^4{\rm Tr}\,[{\mathcal E}(k)]^2$ is given by 

\begin{eqnarray}
\label{eq:3.7}
{\mathcal J}_{e^2}^{(0)}\delta_{e^2}(y) &=& 
u^3\frac{(1-2u+{\hat a}^2u^2)(2-3u-2{\hat a}u^{3/2}+3{\hat a}^2u^2)}{(1+{\hat a}u^{3/2})^2} m_2^2 \partial_{rr} h_{kk}\nonumber\\
&&
- u^5\frac{1-4{\hat a}u^{3/2}+3{\hat a}^2u^2}{(1+{\hat a}u^{3/2})^2} \partial_{\theta\theta} h_{kk}\nonumber\\
&&
 -u^5\frac{(1-3u+2{\hat a}u^{3/2})^2}{(1+{\hat a}u^{3/2})^4(1-2u+{\hat a}^2u^2)}\partial_{\bar\phi\bar\phi} h_{kk}\nonumber\\
&&
+2\frac{u^4}{(1+{\hat a}u^{3/2})^3}[
(1-3u)(2-3u)-3{\hat a}u^{3/2}(1-3u^2)+3{\hat a}^2u^2(3-3u-2u^2)-2{\hat a}^3u^{7/2}(6-5u)\nonumber\\
&&+12{\hat a}^4u^4(1-2u)+12{\hat a}^5u^{11/2}
]m_2 \partial_r h_{kk}\nonumber\\
&&
-2\frac{u^{11/2}}{(1+{\hat a}u^{3/2})^4}[
1-3u-2{\hat a}u^{3/2}(1-6u)+{\hat a}^2u^2(3-17u)+6{\hat a}^3u^{7/2}
] \partial_{r}h_{t\phi} \nonumber\\
&&
-2u^7\frac{(1-3u+2{\hat a}u^{3/2})(1-4{\hat a}u^{3/2}+3{\hat a}^2u^2)}{(1+{\hat a}u^{3/2})^5}\frac{1}{m_2 }(\partial_{r}h_{\phi\phi}-\partial_{\bar \phi} h_{r\phi})\nonumber\\
&&
+2 u^{11/2}\frac{(1-3u+2{\hat a}u^{3/2})(1-4{\hat a}u^{3/2}+3{\hat a}^2u^2)}{(1+{\hat a}u^{3/2})^4} \partial_{\bar \phi}h_{tr}\nonumber\\
&&
+ 2u^7 \frac{(1-2{\hat a}u^{3/2}+{\hat a}^2u^2)^2(1-4{\hat a}u^{3/2}+3{\hat a}^2u^2)^2}{(1-2u+{\hat a}^2u^2)(1+{\hat a}u^{3/2})^4} h_{kk}\nonumber\\
&&
+4u^{15/2}\frac{(1-3u+2{\hat a}u^{3/2})(1-2{\hat a}u^{3/2}+{\hat a}^2u^2)(1-4{\hat a}u^{3/2}+3{\hat a}^2u^2)}{(1-2u+{\hat a}^2u^2)(1+{\hat a}u^{3/2})^5}\frac{1}{m_2}h_{t\phi}\nonumber\\
&&
 -2u^6 \frac{1-2u+{\hat a}^2u^2}{(1+{\hat a}u^{3/2})^4}[
5-18u+18u^2-4{\hat a}u^{3/2}+2{\hat a}^2u^2(6-5u)-12{\hat a}^3u^{7/2}+9{\hat a}^4u^4
] h_{rr}\nonumber\\
&&
 -2u^8\frac{(1-4{\hat a}u^{3/2}+3{\hat a}^2u^2)^2}{(1+{\hat a}u^{3/2})^4}\frac{1}{m_2^2}h_{\theta\theta}\nonumber\\
&&
 +2u^8\frac{(1-3u+2{\hat a}u^{3/2})[1-u+2{\hat a}u^{3/2}(1-2u)+2{\hat a}^2u^3](1-4{\hat a}u^{3/2}+3{\hat a}^2u^2)}{(1-2u+{\hat a}^2u^2)(1+{\hat a}u^{3/2})^6}\frac{1}{m_2^2}h_{\phi\phi}\,.
\end{eqnarray}

\subsection{Quadratic magnetic-type invariant ${\rm Tr}\,[{\mathcal B}(k)]^2$}

The $O(q)$ perturbation to $m_2^4{\rm Tr}\,[{\mathcal B}(k)]^2$ is given by 
\begin{eqnarray}
{\mathcal J}_{b^2}^{(0)}\delta_{b^2}(y)
&=&
3u^4\frac{1-{\hat a}u^{1/2}}{(1+{\hat a}u^{3/2})^3}(1-2{\hat a}u^{3/2}+{\hat a}^2u^2)(1-2u+{\hat a}^2u^2)  m_2^2 \partial_{rr}  h_{kk}\nonumber\\
&&
- 3u^6\frac{1-{\hat a}u^{1/2}}{(1+{\hat a}u^{3/2})^3}(1-2{\hat a}u^{3/2}+{\hat a}^2u^2)\partial_{\theta\theta}h_{kk}\nonumber\\
&&  
+ 3u^{9/2}\frac{1-{\hat a}u^{1/2}}{(1+{\hat a}u^{3/2})^4}(1-3u+2{\hat a}u^{3/2})(1-2u+{\hat a}^2u^2)\left(m_2 \partial_{rr}  h_{t\phi}+\frac{u^{3/2}}{1+{\hat a}u^{3/2}}\partial_{rr}   h_{\phi\phi}\right)\nonumber\\
&&
 -3u^{13/2}\frac{1-{\hat a}u^{1/2}}{(1+{\hat a}u^{3/2})^4}(1-3u+2{\hat a}u^{3/2})\frac{1}{m_2^2}\left(m_2\partial_{\theta\theta} h_{t\phi}+\frac{u^{3/2}}{1+{\hat a}u^{3/2}}\partial_{\theta\theta}  h_{\phi\phi}\right)\nonumber\\
&&
- 3u^{9/2}\frac{1-{\hat a}u^{1/2}}{(1+{\hat a}u^{3/2})^4}(1-3u+2{\hat a}u^{3/2})(1-2u+{\hat a}^2u^2) \left(m_2\partial_{r\phi}  h_{tr}+\frac{u^{3/2}}{1+{\hat a}u^{3/2}}\partial_{r\phi}   h_{r\phi}\right)\nonumber\\
&&
+3u^{13/2}\frac{1-{\hat a}u^{1/2}}{(1+{\hat a}u^{3/2})^4}(1-3u+2{\hat a}u^{3/2})\frac{1}{m_2^2}\left(m_2\partial_{\theta\phi}  h_{t\theta}+\frac{u^{3/2}}{1+{\hat a}u^{3/2}}\partial_{\theta\phi} h_{\theta\phi}\right)\nonumber\\
&&
- 3u^5\frac{1-{\hat a}u^{1/2}}{(1+{\hat a}u^{3/2})^4}[
-3(1-3u)+2{\hat a}u^{1/2}(4-15u+12u^2)+{\hat a}^2u^2(1-3u)(13-8u)\nonumber\\
&&
+2{\hat a}^3u^{5/2}(5-3u-13u^2)
+2{\hat a}^4u^4(11-5u)+12{\hat a}^5u^{11/2}
] m_2\partial_r h_{kk}\nonumber\\
&&
- 6u^{11/2}\frac{1-{\hat a}u^{1/2}}{(1+{\hat a}u^{3/2})^5}[
(1-u)(1-3u)+3{\hat a}u^{3/2}(1-3u+4u^2)+{\hat a}^2u^2(1-11u^2)\nonumber\\
&&
+{\hat a}^3u^{7/2}(3-u)+2{\hat a}^4u^5
]\partial_r h_{t\phi}\nonumber\\
&&
-3u^7\frac{1-{\hat a}u^{1/2}}{(1+{\hat a}u^{3/2})^6}(1-3u+2{\hat a}u^{3/2})[
3-5u+4{\hat a}u^{3/2}(1-2u)+2{\hat a}^2u^2(1+u)+2{\hat a}^3u^{7/2}
]\frac{1}{m_2}\partial_r h_{\phi\phi}\nonumber\\
&&
-3u^5\frac{1-{\hat a}u^{1/2}}{(1+{\hat a}u^{3/2})^4}(1-3u+2{\hat a}u^{3/2})(1-2u+{\hat a}^2u^2)^2 m_2 \partial_r h_{rr}\nonumber\\
&&
 -3u^7\frac{1-{\hat a}u^{1/2}}{(1+{\hat a}u^{3/2})^4}(1-3u+2{\hat a}u^{3/2})(1-2u+{\hat a}^2u^2)\frac{1}{m_2}\left(\partial_r h_{\theta\theta}-2\partial_\theta h_{r\theta}\right)\nonumber\\
&&
+6u^{11/2}\frac{1-{\hat a}u^{1/2}}{(1+{\hat a}u^{3/2})^4}(1-3u+2{\hat a}u^{3/2})(1-2u+{\hat a}^2u^2)\frac{1}{m_2}\left(m_2 \partial_\phi   h_{tr}+\frac{u^{3/2}}{1+{\hat a}u^{3/2}}\partial_\phi h_{r\phi}\right)\nonumber\\
&&
 + 6u^6\frac{1-{\hat a}u^{1/2}}{(1+{\hat a}u^{3/2})^4}(1-2u+{\hat a}^2u^2)[
1-9u+15u^2+{\hat a}u^{3/2}(5-12u)+{\hat a}^2u^2(2-9u)+7{\hat a}^3u^{7/2}
]  h_{rr}\nonumber\\
&&
+6u^8\frac{1-{\hat a}u^{1/2}}{(1+{\hat a}u^{3/2})^6}\frac{1-3u+2{\hat a}u^{3/2}}{1-2u+{\hat a}^2u^2}[
1-3u+3u^2+{\hat a}u^{3/2}(1-2u)(5-6u)+7{\hat a}^2u^3(1-2u)\nonumber\\
&&+{\hat a}^3u^{7/2}(3-4u)+6{\hat a}^4u^5
]\frac{1}{m_2^2} h_{\phi\phi}\nonumber\\
&&
-18u^9\frac{(1-{\hat a}u^{1/2})^2}{(1+{\hat a}u^{3/2})^4}(1-2u+{\hat a}^2u^2)\frac{1}{m_2^2} h_{\theta\theta}\nonumber\\
&&
+12u^{15/2}\frac{1-{\hat a}u^{1/2}}{(1+{\hat a}u^{3/2})^5(1-2u+{\hat a}^2u^2)}[
(1-3u)(2-3u)+6{\hat a}u^{5/2}(2-3u)+{\hat a}^2u^2(5-30u+33u^2)\nonumber\\
&&+8{\hat a}^3u^{7/2}(1-u)+{\hat a}^4u^4(3-13u)+6{\hat a}^5u^{11/2}
]\frac{1}{m_2} h_{t\phi}\nonumber\\
&&
+6u^{15/2}\frac{1-{\hat a}u^{1/2}}{(1+{\hat a}u^{3/2})^4(1-2u+{\hat a}^2u^2)}[
u^{1/2}+{\hat a}(3-18u+20u^2)+2{\hat a}^2u^{3/2}(3-3u-4u^2)\nonumber\\
&&+2{\hat a}^3u^2(3-14u+12u^2)+{\hat a}^4u^{7/2}(12-11u)+{\hat a}^5u^4(3-10u)+6{\hat a}^6u^{11/2}
] h_{kk}\,.
\end{eqnarray}

\subsection{Cubic electric-type invariant ${\rm Tr}\,[{\mathcal E}(k)]^3$}

The $O(q)$ perturbation to $m_2^6{\rm Tr}\,[{\mathcal E}(k)]^3$ is given by
\begin{eqnarray}
{\mathcal J}_{e^3}^{(0)}\delta_{e^3}(y)&=& 
-\frac32\frac{u^6}{(1+{\hat a}u^{3/2})^4}(1-2u+{\hat a}^2u^2)(2-3u+2{\hat a}u^{3/2}+3{\hat a}^2u^2)^2 m_2^2 \partial_{rr} h_{kk}\nonumber\\
&&
-\frac32\frac{u^8}{(1+{\hat a}u^{3/2})^4}(1-4{\hat a}u^{3/2}+3{\hat a}^2u^2)\partial_{\theta\theta} h_{kk}\nonumber\\
&&
-\frac32\frac{u^8}{(1+{\hat a}u^{3/2})^6}\frac{(1-3u+2{\hat a}u^{3/2})^3}{1-2u+{\hat a}^2u^2}\partial_{\bar \phi\bar \phi}h_{kk}\nonumber\\
&&
-3\frac{u^7}{(1+{\hat a}u^{3/2})^5}(1-3u+2{\hat a}u^{3/2})[
(2-3u)^2-3{\hat a}u^{3/2}(1-u)(5-3u)+18{\hat a}^2u^2(1-u+u^2)\nonumber\\
&&-2{\hat a}^3u^{7/2}(15-u)+3{\hat a}^4u^4(7-8u)+12{\hat a}^5u^{11/2}
]m_2\partial_r h_{kk}\nonumber\\
&&
 +3\frac{u^{17/2}}{(1+{\hat a}u^{3/2})^6}(1-3u+2{\hat a}u^{3/2})[
5-18u+18u^2-4{\hat a}u^{3/2}+2{\hat a}^2u^2(6-5u)\nonumber\\
&&-12{\hat a}^3u^{7/2}+9{\hat a}^4u^4]
\frac{1}{m_2}\left[
m_2\left(\partial_r h_{t\phi}-\partial_{\bar \phi}h_{tr}\right)
+\frac{u^{3/2}}{1+{\hat a}u^{3/2}}\left(\partial_r  h_{\phi\phi}-\partial_{\bar \phi} h_{r\phi}\right)
\right]\nonumber\\
&&
 -3\frac{u^{10}}{(1+{\hat a}u^{3/2})^6}\frac{(1-2{\hat a}u^{3/2}+{\hat a}^2u^2)^2}{1-2u+{\hat a}^2u^2}[
5-18u+18u^2-4{\hat a}u^{3/2}+2{\hat a}^2u^2(6-5u)-12{\hat a}^3u^{7/2}+9{\hat a}^4u^4
]h_{kk}\nonumber\\
&&
 -3\frac{u^{11}}{(1+{\hat a}u^{3/2})^8}\frac{1-3u+2{\hat a}u^{3/2}}{1-2u+{\hat a}^2u^2}[1-u+2{\hat a}u^{3/2}(1-2u)+2{\hat a}^2u^3][
5-18u+18u^2-4{\hat a}u^{3/2}\nonumber\\
&&+2{\hat a}^2u^2(6-5u)-12{\hat a}^3u^{7/2}+9{\hat a}^4u^4
]\frac{1}{m_2^2} h_{\phi\phi}\nonumber\\
&&
-6\frac{u^{21/2}}{(1+{\hat a}u^{3/2})^7}\frac{1-3u+2{\hat a}u^{3/2}}{1-2u+{\hat a}^2u^2}(1-2{\hat a}u^{3/2}+{\hat a}^2u^2)[
5-18u+18u^2-4{\hat a}u^{3/2}+2{\hat a}^2u^2(6-5u)\nonumber\\
&&-12{\hat a}^3u^{7/2}+9{\hat a}^4u^4
]\frac{1}{m_2} h_{t\phi}\nonumber\\
&&
 -3\frac{u^{11}}{(1+{\hat a}u^{3/2})^6}(1-4{\hat a}u^{3/2}+3{\hat a}^2u^2)^3\frac{1}{m_2^2}h_{\theta \theta}\nonumber\\
&&
+3\frac{u^9}{(1+{\hat a}u^{3/2})^6}(1-4{\hat a}u^{3/2}+3{\hat a}^2u^2)(1-2u+{\hat a}^2u^2)[
7-27u+27u^2-2{\hat a}u^{3/2}+{\hat a}^2u^2(15-23u)\nonumber\\
&&-6{\hat a}^3u^{7/2}+9{\hat a}^4u^4
] h_{rr}
\,.
\end{eqnarray}

\end{widetext}

\section*{Acknowledgments}

The authors thank T. Damour for useful discussions.
D.B. thanks the Naples Section of the Italian Istituto Nazionale di Fisica Nucleare (INFN) and the International Center for Relativistic Astrophysics Network (ICRANet) for partial support.

\end{document}